\documentclass[conference]{IEEEtran}
\IEEEoverridecommandlockouts
\def\BibTeX{{\rm B\kern-.05em{\sc i\kern-.025em b}\kern-.08em
    T\kern-.1667em\lower.7ex\hbox{E}\kern-.125emX}}

\usepackage{pgfplots}
\usepackage{stfloats}
\usepackage{colortbl}
\usepackage{multirow}
\usepackage{hhline}
\usepackage{cite}
\usepackage{amsmath,amssymb,amsfonts}
\usepackage{verbatim}
\usepackage[linesnumbered,ruled,vlined]{algorithm2e}
\usepackage{multirow}
\usepackage{graphicx}
\usepackage{booktabs}
\usepackage{textcomp}
\usepackage{xcolor}
\usepackage{stackengine}
\usepackage{float}
\usepackage{subfig}
\usepackage{multicol}
\usepackage{hyperref}
\usepackage{makecell}
\usepackage{threeparttable}
\usepackage[inkscapeformat=png]{svg}
\usepackage[a4paper, total={184mm,239mm}]
{geometry}
\def\BibTeX{{\rm B\kern-.05em{\sc i\kern-.025em b}\kern-.08emT\kern-.1667em\lower.7ex\hbox{E}\kern-.125emX}}

\SetKwInput{KwInput}{Input}
\SetKwInput{KwOutput}{Output}

\newcommand{\name}{AIA}

\usepackage{soul}

\begin{document}

\title{{\name{}: A 16nm Multicore SoC for Approximate Inference Acceleration Exploiting Non-normalized Knuth-Yao Sampling and Inter-Core Register Sharing}\\
}
\author{ 
\IEEEauthorblockN{ Shirui~Zhao\IEEEauthorrefmark{1},
Nimish~Shah\IEEEauthorrefmark{1}, 
Wannes~Meert\IEEEauthorrefmark{2}, and~Marian~Verhelst\IEEEauthorrefmark{1}}

\IEEEauthorblockA{
\IEEEauthorrefmark{1}MICAS-ESAT, KU Leuven,
\IEEEauthorrefmark{2}DTAI, KU Leuven
}
\IEEEauthorblockA{\{shirui.zhao, marian.verhelst\}@kuleuven.be} 
}

\maketitle

\begin{abstract}

Probabilistic graphical models (PMs) are popular to empower machine learning with the ability of reasoning and decision-making. 
To perform approximate inference in PMs, sampling-based Markov Chain Monte Carlo (MCMC) algorithms are commonly employed. Unfortunately, MCMC is compute-intensive and hard to run in parallel, resulting in inefficient execution on modern CPU/GPU platforms.
This paper proposes \name{}, an Approximate Inference Accelerator designed to empower decision-making and reasoning at the edge.
\name{} consists of a RISC-V host, and a 2D mesh of 16 customized RISC-V cores optimized to efficiently support PM inference, each featuring (i) a novel non-normalized Knuth-Yao sampler and interpolation unit; and (ii) core-to-core direct data access via the register file, which provides solutions for compute-intensive operations. To fully exploit the parallel potential of Markov Chain Monte Carlo (MCMC) algorithms, a customized compiler chain has been developed for effective spatial mapping and scheduling on the chip.
\name{} can generate 1277 MSample/s at 0.9V and 20 GSamples/s/W at 0.7V which is up to 2$\times$ faster and 1.45x more energy efficient compared to the previous state-of-the-art Markov Random Field (MRF) accelerator.
We further map Bayesian Networks benchmark onto \name{} to show the flexibility of our design.
\end{abstract}

\begin{IEEEkeywords}
Probabilistic graphical models, Bayesian inference, Approximate inference, MCMC, Knuth-Yao sampling, RISC-V
\end{IEEEkeywords}

\section{Introduction}
\begin{figure}[!t]
    \centering
    \includegraphics[trim={0cm 0cm 0cm 0cm} , clip, width=\columnwidth]{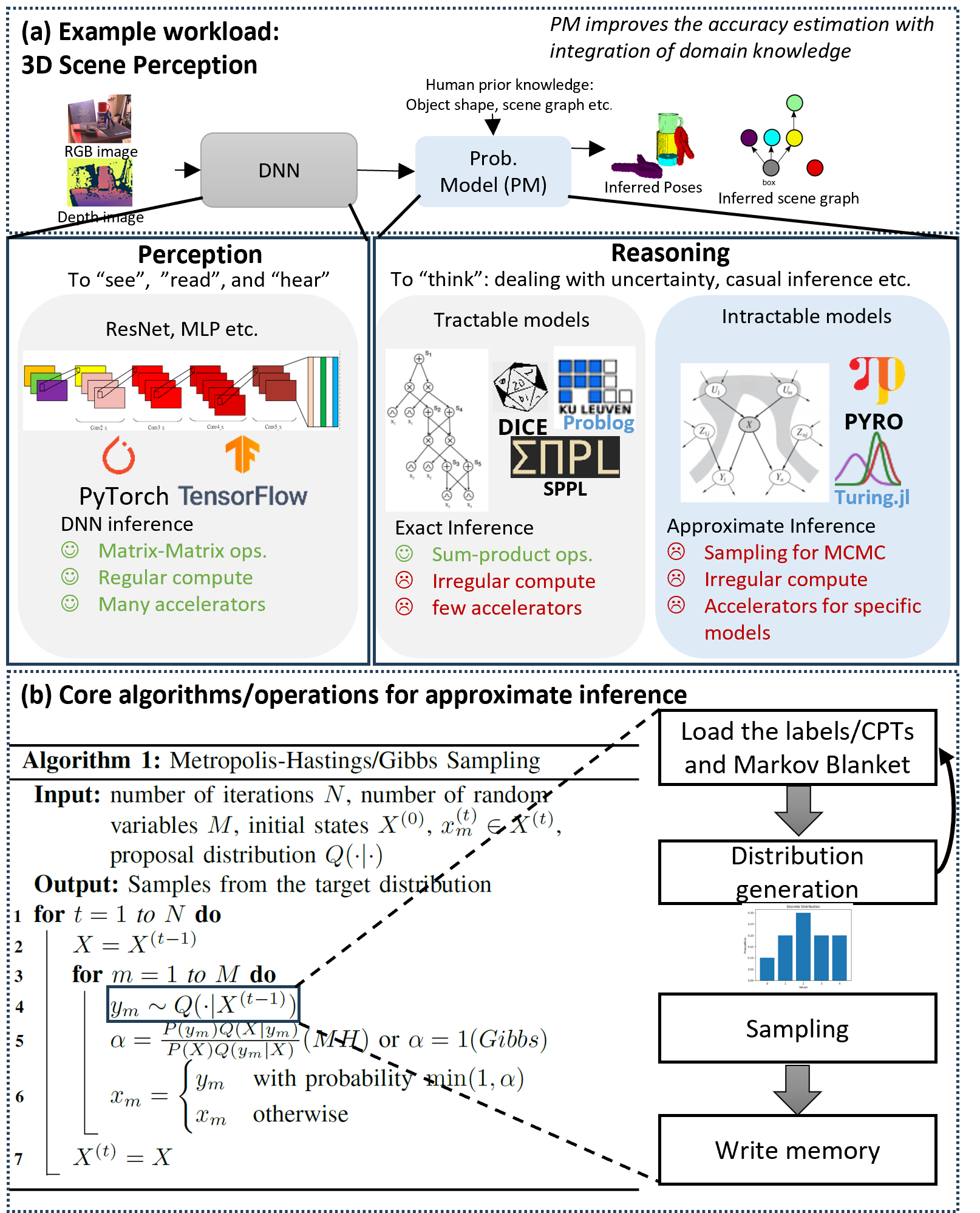}
    \caption{Deep learning v.s. probabilistic machine learning.}
    \label{fig:ml_overview}
\end{figure}
Edge intelligence requires both perception and reasoning capabilities, also denoted as “hybrid AI”. As shown in Fig. \ref{fig:ml_overview}, Deep neural networks (DNN) have brought enormous successes for perception tasks, yet proven to be ill-suited for reasoning tasks\cite{bdl}.  A more potent framework for addressing a spectrum of complex decision-making tasks is probabilistic graphical models (PM)\cite{agrum}, which allows for integrating human knowledge into the models and handling reasoning with uncertainties. For example, in a 3D scene application (Fig. \ref{fig:ml_overview})(a), the human prior knowledge about the size and relationship between objects in the 3D image facilitates a more accurate scene understanding\cite{3dp3}. In other fields such as finance\cite{bn_finance}, pharmaceuticals \cite{bn_medicine}, etc., which require explainable "white-box" models, PMs are also preferred over their DNN counterparts for certain applications.

The PMs can be represented with (un)directed graphs, with nodes representing random variables (RVs) and edges indicating relationships between RVs.
However, when scaling up the complexity of the graphs, the computing quickly becomes intractable. instead of exact inference\cite{dpu}, MCMC techniques offer a scalable solution by generating samples from the posterior distribution\cite{lp_mcmc}.
The core operation of MCMC is illustrated in Fig. \ref{fig:ml_overview}(b). The parallelism is hard to utilize as it requires processes N iterations to update all RVs in the graph until convergence to the desired target distributions.
Line 4 denotes the update of one RV, which needs 
\textcircled{\raisebox{-0.9pt}{1}} irregular data access for its neighbors' RVs in the Markov Blanket (MB),
\textcircled{\raisebox{-0.9pt}{2}} perform complex computing (exp, log, etc) to generate distribution for this RV, and 
\textcircled{\raisebox{-0.9pt}{3}} sample from the distribution. This results in very poor computing efficiency on CPU and GPU.
Moreover, State-of-the-art neural network accelerators also do not match with the irregular graph-based traversal and computational kernels. 
As a result, specific ASIC accelerators have been developed targeting the inference over PMs. The ASIC PM accelerators available in the SotA are highly customized, lacking software programmability to support multiple types of PMs\cite{pgma}\cite{spu} or only support exact inference\cite{dpu} which is not suitable for large PMs.

This paper presents \name{}, the first silicon implementation of a multi-core programmable approximate inference accelerator with RISC-V ISA extensions. \name{} boosts the efficiency of MCMC for various types of probabilistic graphical models, including Markov Random Field (MRF) and Bayesian Networks (BN).
\section{\name{} Hardware Design}
\begin{figure}[!t]
    \centering
    \includegraphics[trim={0cm 0cm 0cm 0cm} , clip, width=0.8\columnwidth]{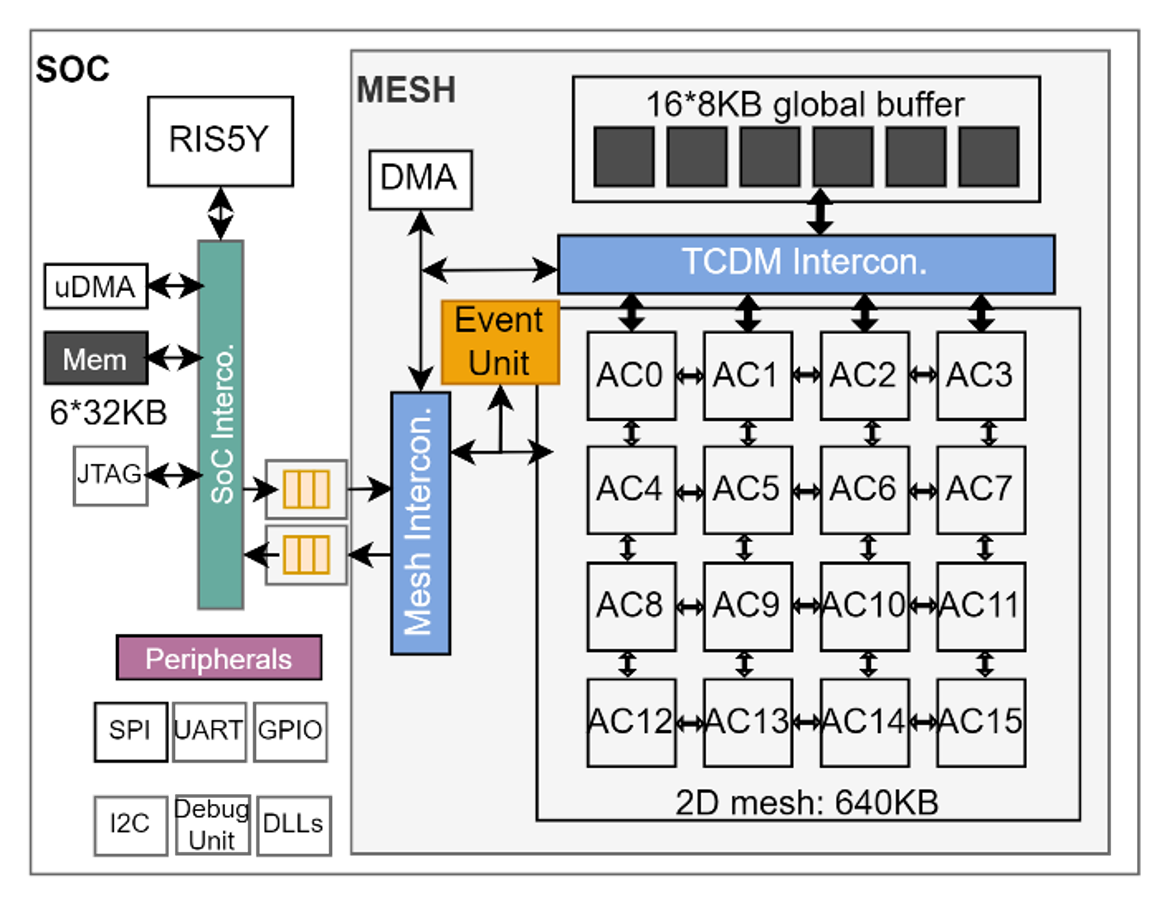}
    \caption{The \name{} architecture with 16 accelerator cores(AC)}
    \label{fig:top_arch}
\end{figure}
Fig. \ref{fig:top_arch} shows the 16nm \name{} SoC with a RIS5Y host core\cite{pulp} and a 4x4 mesh of customized accelerator cores (AC). The host core is responsible for connecting the accelerator to the outside over GPIO, SPI, I2C, Hyperram, and UART interface, which contains 192KB SoC memory for data buffering. The accelerator mesh sits in a separate clock domain and contains 4x4 ACs, an event unit for AC synchronization, and a tightly-coupled 128KB global buffer accessible by the top 4 ACs. 

\begin{figure}[!t]
    \centering
    \includegraphics[trim={0cm 0cm 0cm 0cm} , clip, width=0.85\columnwidth]{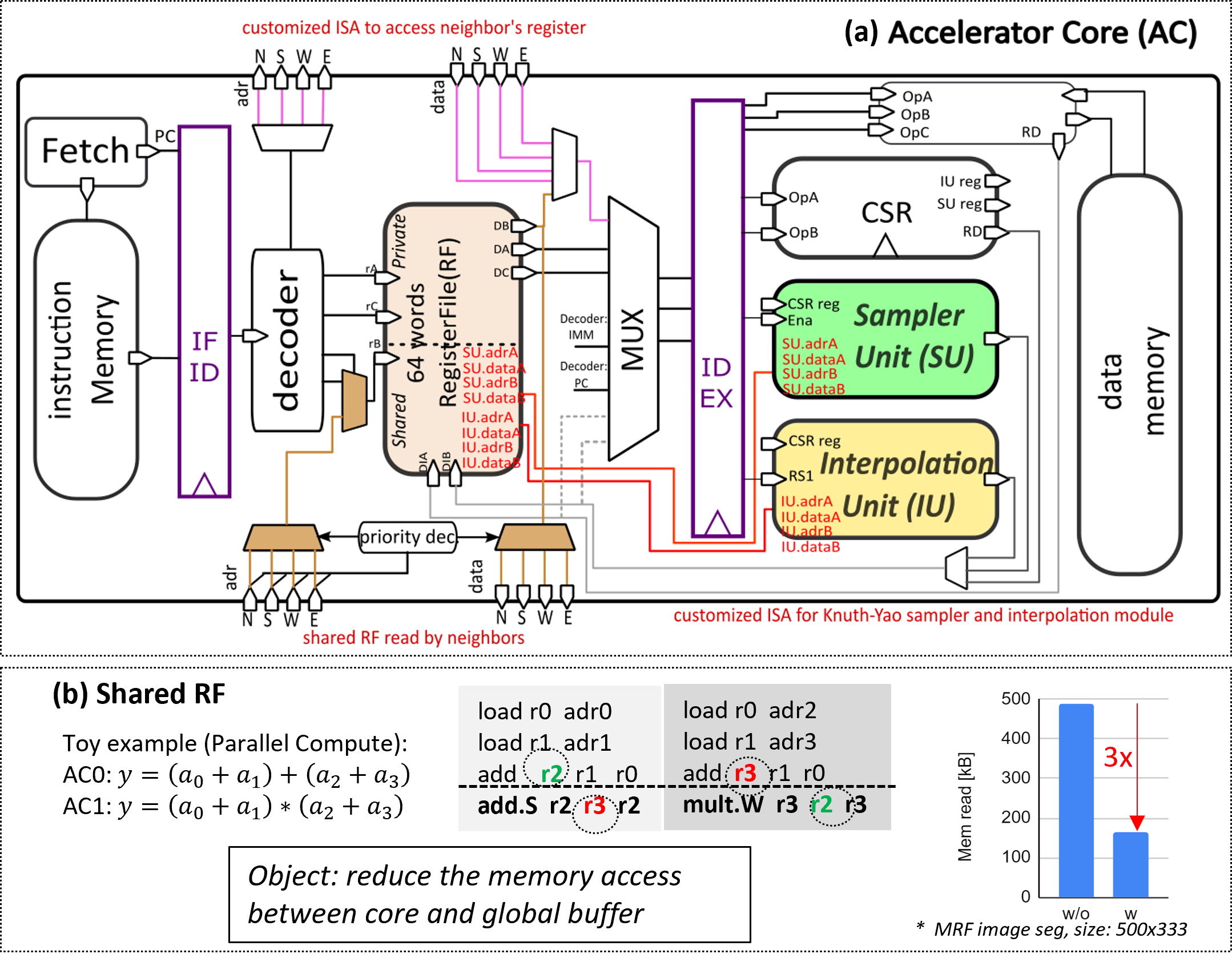}
    \caption{(a) The AC micro-architecture; (b) Example of using inter-core register sharing ISA.}
    \label{fig:ac_arch}
\end{figure}

As shown in Fig. \ref{fig:ac_arch}, each AC is a customized RISC-V core \cite{pulp} with an 8KB instruction and a 32 KB data scratchpad, extended with A) neighboring core register access; and B) custom instructions for sampling and interpolation.

\begin{figure}[!t]
    \centering
    \includegraphics[trim={0cm 0cm 0cm 0cm} , clip, width=\columnwidth]{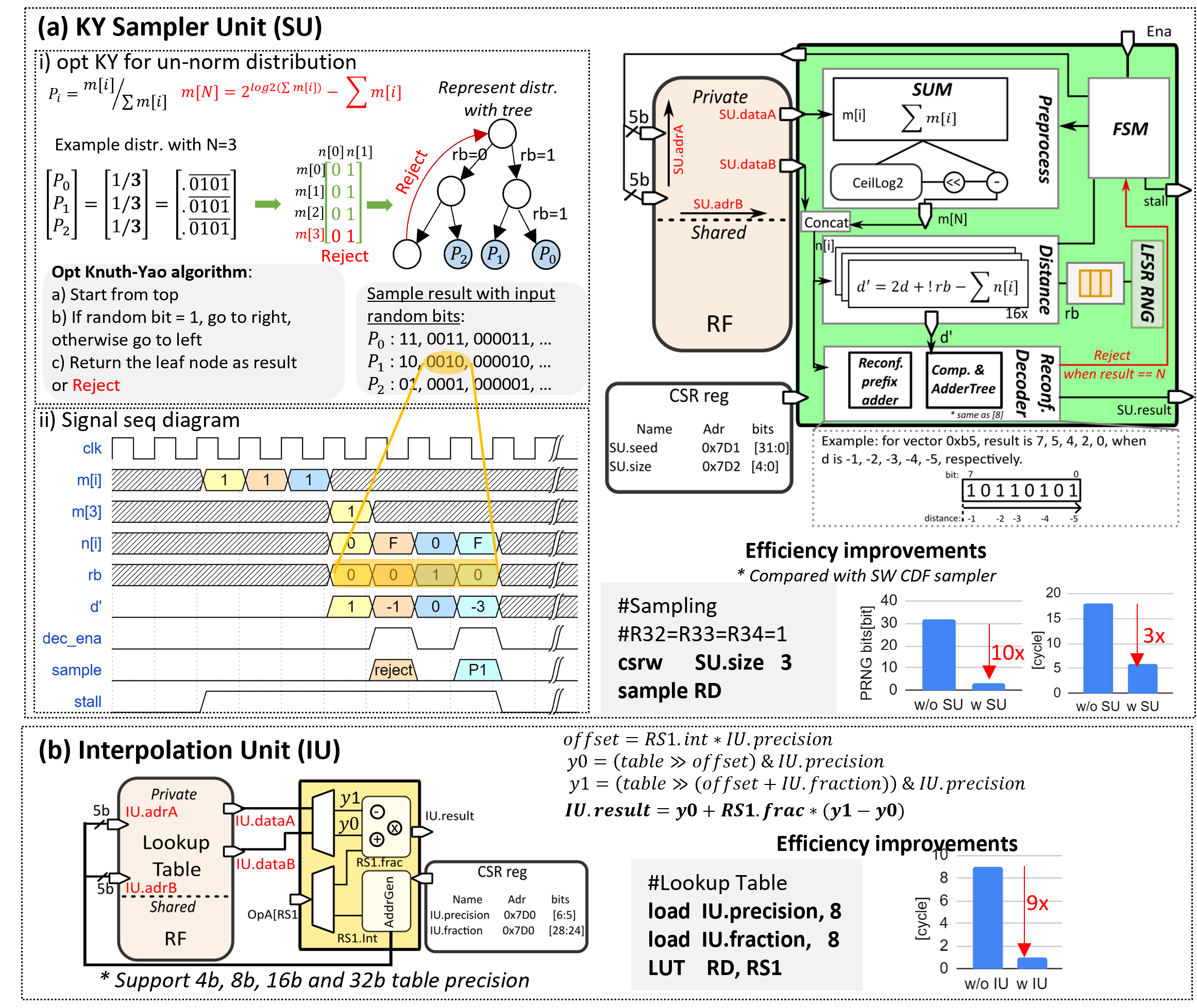}
    \caption{(a) SU for non-normalized integer distributions; (b) IU for LUT function of complex operations.}
    \label{fig:ky_iu}
\end{figure}

\subsection{Inter-core register sharing}
MCMC frequently requires access to its neighbor RVs, hence benefiting from a 2D mesh architecture for data communication between processing cores dealing with different nodes in parallel. 
Each AC contains a register file (RF) with a size of 64 words. To enable this sharing without costly memory accesses, the ACs have been extended with single-cycle read access capabilities into 32 shared registers of their N-E-S-W neighbors, as shown in Fig. \ref{fig:ac_arch}(a). 
A simple example in Fig. \ref{fig:ac_arch}(b) shows the benefit of using the RF sharing feature. It can avoid the global buffer operation instruction after computing. Given the MRF image segmentation task, we can reduce 3x memory read overhead via this approach.

\subsection{ISA extension for sampling and interpolation}
Sampling from arbitrary discrete distributions is the key operation in the MCMC algorithm. In existing works\cite{agrum}\cite{pgma}\cite{spu}, the traditional cumulative distribution function(CDF) samplers are widely used, which shows low performance compared to the Knuth-Yao (KY) algorithm\cite{sz_date}\cite{fldr}. 
We propose a novel KY hardware sampler unit(SU) with rejection feature. The algorithm is described in Fig. \ref{fig:ky_iu}(a). It starts from the root node and traverses the tree using random bits from Linear Feedback Shift Registers (LFSR). When it reaches a leaf node, it returns the corresponding sample result. Unlike the initial KY sampler, the novel one will restart when reaching the rejected item. This method effectively reduces the need for extensive cumulative and normalization operations while simultaneously minimizing the memory footprint required for the distribution.

The SU is integrated into the AC's decode and execution stage. After distribution generation, the non-normalized distribution is stored in RF in a row-wise way. The KY sampling algorithm requires column-wise access to the probabilities for tree search after computing the rejected item. To this end, we add 2 additional read ports to the RF (a row-wise and a column-wise one). The sampling instruction requires a variable number of clock cycles, during which the rest of the processor pipeline is automatically stalled. An example of sampling from the distribution with
$P_x = \frac{1}{3}, \quad \text{for} \quad x \in [0, 2]$ is shown in Fig. \ref{fig:ky_iu}(a).
It consumes an average of 3 random bits per sample, resulting in a 3x reduction in runtime compared to the CDF sampler.

A second ISA extension is introduced for look-up-table (LUT)-based interpolation unit (IU). MCMC requires repetitively for complex operations (exp, log, etc.) in the distribution generation stage of all RVs. This single-cycle datapath extension also requires two additional RF read ports, directly accessible in the EX stage based on the interpolation index fetched from the register file through RS1. As shown in Fig. \ref{fig:ky_iu}(b), the $offset$ and $IU.address$ are generated according to the integer part of the input data. Following the formula in the figure, the linear interpolation result is calculated in one cycle. Compared to memory-based LUT, it can save 9x runtime.

The customized ISA are defined in Table \ref{tab:isa}. As the SU/IU requires an enlarged RF address space for its private RF and neighbors' shared RF, the address index width is extended by 3 bits for the customized arithmetic R-type operations. Also, we added two instructions to trigger the IU and SU.
\begin{table}[!t]
\centering
\caption{Customized RISC-V ISA}
\begin{tabular}{lccc}
\toprule
Function & \makecell[c]{Func8\\31:29} & \makecell[c]{Func7\\28:25} & \makecell[c]{Func3\\14:12} \\ \midrule
private RF & 0 & 0-9 & {RD[5], RS1[5], RS2[5]}  \\
shared RF & 1 & 0-9 & 0:left shared RF, etc.  \\ 
\midrule
SU & 2 & 0x0 & 0 \\
\midrule
IU & 3 & 0x0 & 0 \\
\bottomrule
\end{tabular}%
\label{tab:isa}
\end{table}

\section{Customized Compiler}
\begin{figure}[!t]
    \centering
    \includegraphics[trim={0cm 0cm 0cm 0cm} , clip, width=0.82\columnwidth]{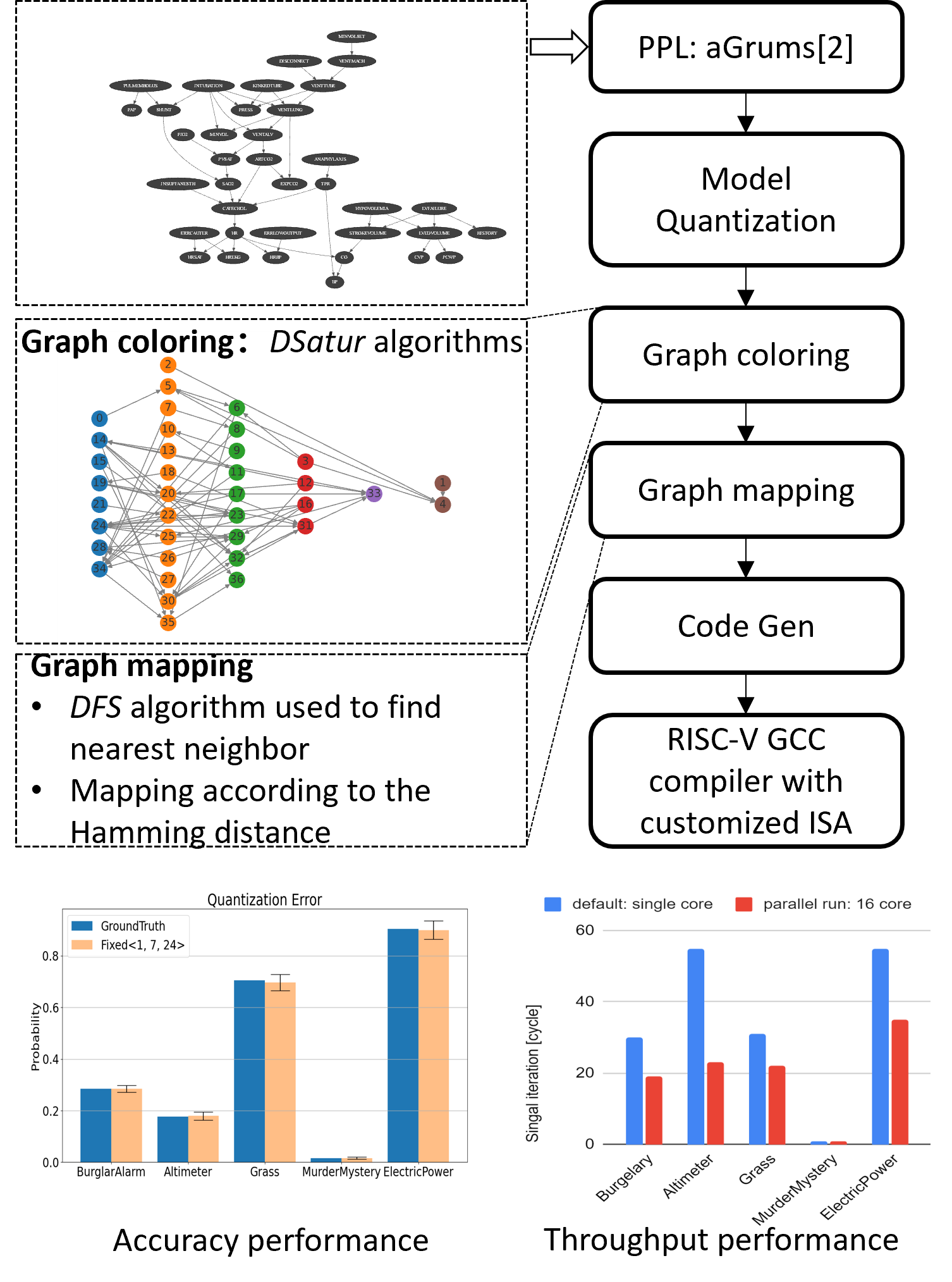}
    \caption{Customized compiler chain for \name{}.}
    \label{fig:compiler}
\end{figure}

A custom compiler chain  (Fig. \ref{fig:compiler}) is developed for \name{}, to enable rapid mapping of MCMC algorithms for a wide variety of PMs.

The PMs can be described using a probabilistic programming language (PPL): aGrums\cite{agrum}, then subsequently go through a fixed-point model quantization stage \cite{lp_mcmc} with negligible accuracy loss (see Fig. \ref{fig:compiler}), and a graph coloring stage. Graph coloring splits all nodes (variables) of the models into sets of conditionally independent variables (colors), which can be updated in parallel. E.g. for MRF, block Gibbs sampling requires only two colors in a checkerboard pattern\cite{pgma}\cite{spu}. 
Irregular workloads, such as Bayesian Networks \cite{bn_finance}\cite{bn_medicine}, typically require more sets.
We utilize aGrum\cite{agrum} to convert the Directed Acyclic Graph (DAG) to the factor graph and exploit the heuristic DSatur\cite{networkx} to color the graph. 

Afterward, mutually independent nodes are mapped onto the 16 parallel AC cores, using the information from the graph coloring stage, with a heuristic that maximizes the parallelism and minimizes the communication distance between nodes that have information exchange. Finally, using the custom RISC-V GCC compiler, a binary is generated for each core to execute the respective probability generation and node sampling.

\section{Measurement Results}
\begin{figure}
    \centering
    \includegraphics[trim={0cm 0cm 0cm 0cm} , clip, width=0.82\columnwidth]{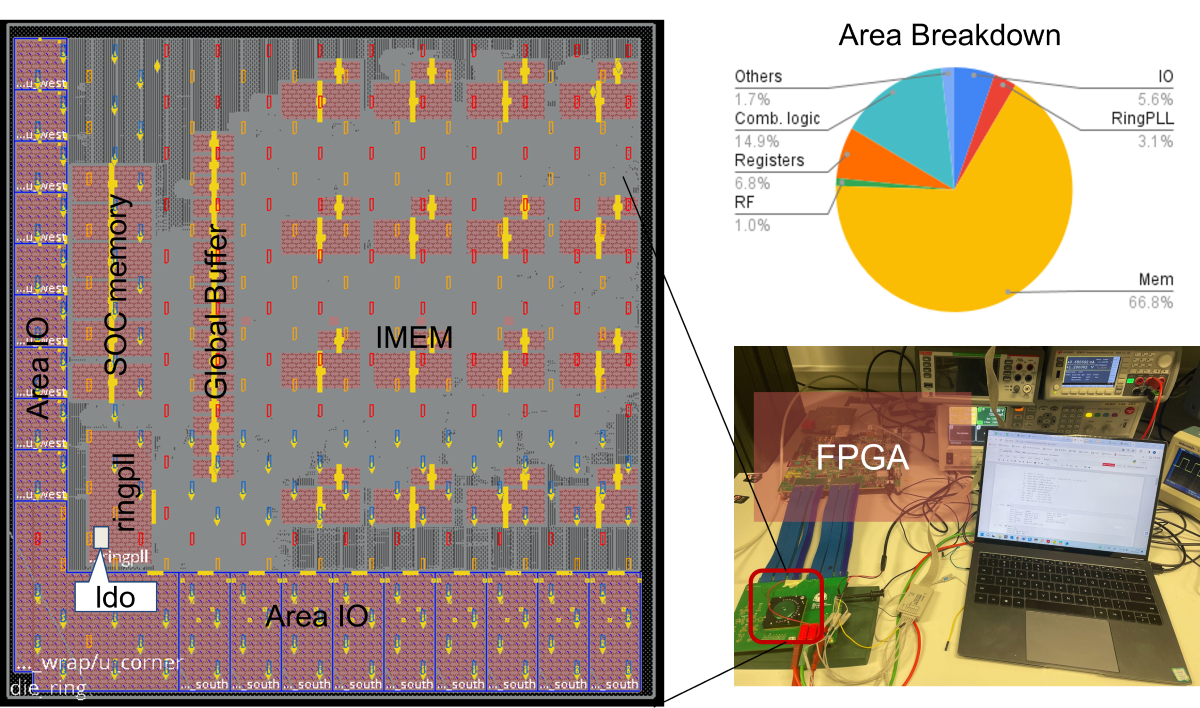}
    \caption{Chip post layout photo, area breakdown and test environment.}
    \label{fig:test_photo}
\end{figure}

\name{} is implemented in Intel 16nm technology. Fig. \ref{fig:test_photo} shows the die photo with area of $4mm^2$ and test environment where the memory occupied $66.8\%$ in the area breakdown.

\begin{figure}[!t]
    \centering
    \includegraphics[trim={0cm 0cm 0cm 0cm} , clip, width=0.9\columnwidth]{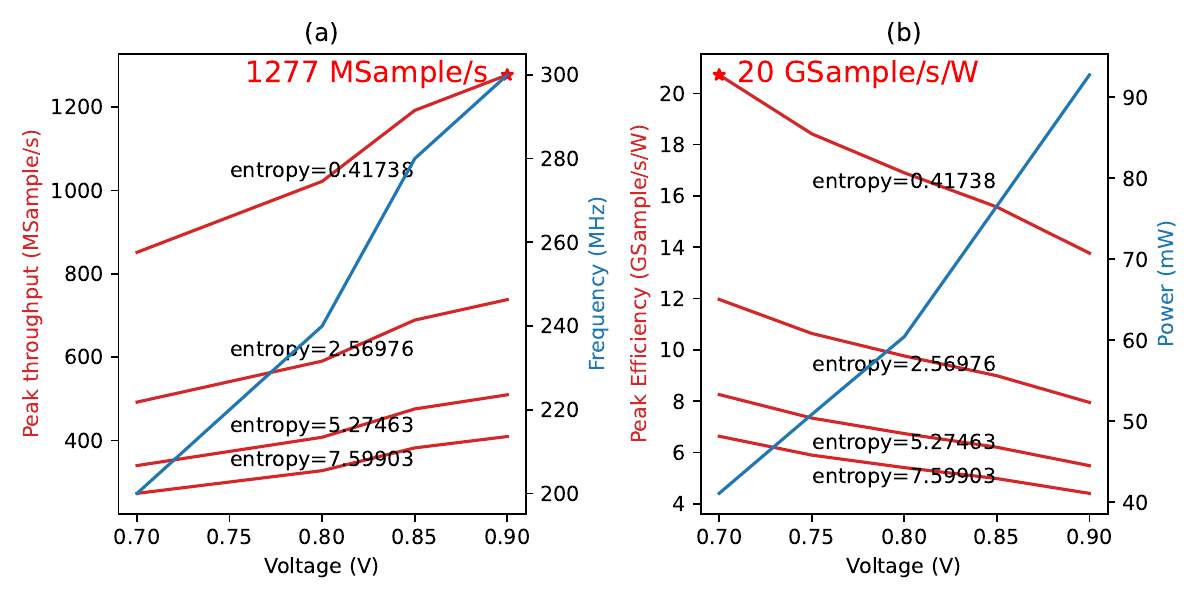}
    \caption{Peak performance scaling with frequency and voltage. $\star$ denotes the peak throughput and energy efficiency performance.
    }
    \label{fig:shmoo}
\end{figure}

Fig. \ref{fig:shmoo} shows the chip's Schmoo plot which is run on sampling from the distributions with different entropies. The peak throughput performance is attained at 300MHz with a voltage of 0.9V, while the highest energy efficiency is reached at a voltage of 0.7V.

\begin{figure}[!t]
    \centering
    \includegraphics[trim={0cm 0cm 0cm 0cm} , clip, width=0.82\columnwidth]{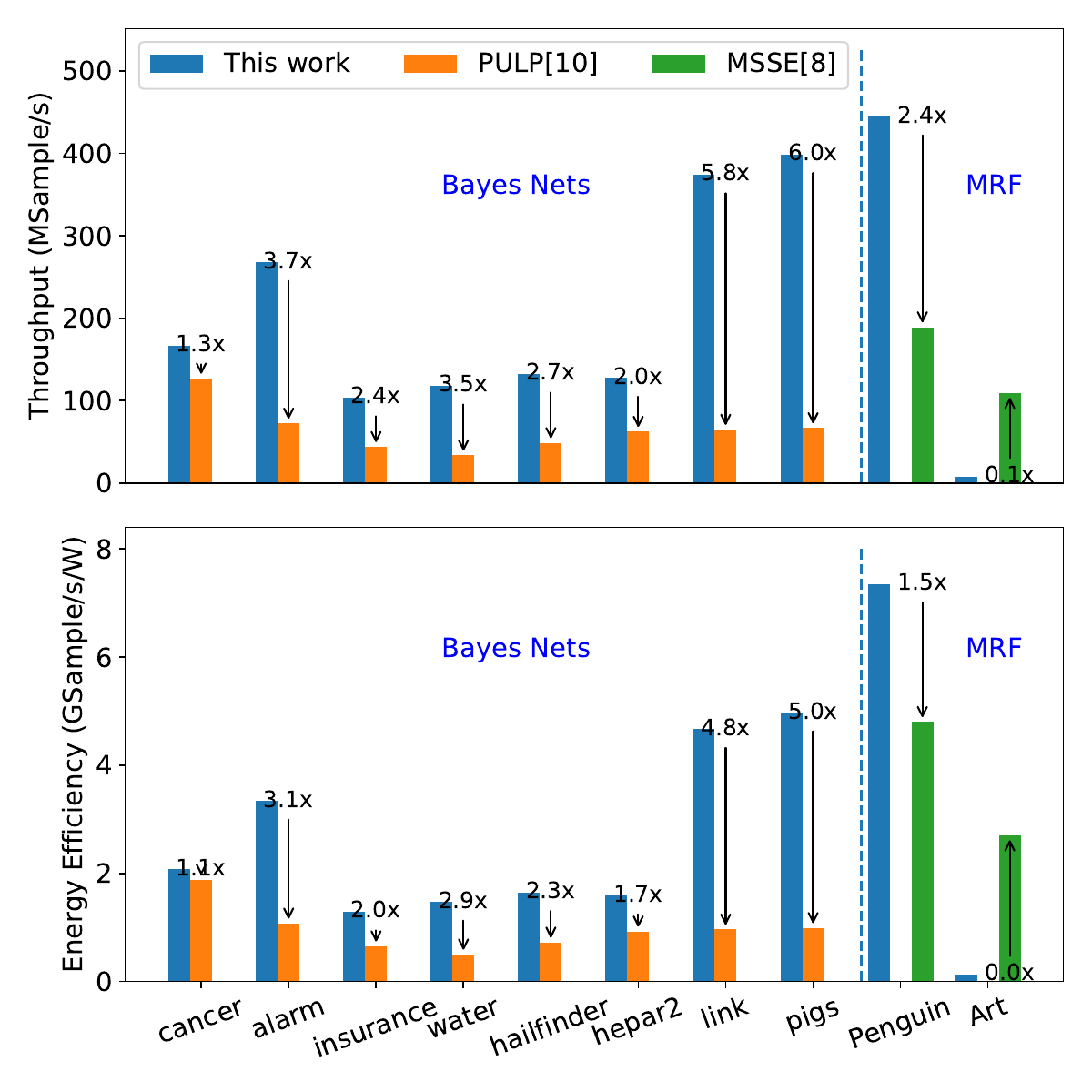}
    \caption{Throughput and energy performance compassion on real-world applications.}
    \label{fig:application_result}
\end{figure}

Next, Two types of end-to-end applications are mapped to the chip for SotA benchmarking: (a) Bayes Nets\cite{bnrepository}; (b) two MRF datasets from \cite{pgma}: image segmentation for Penguin (image size of 500x333, label size of 2); stereo match for Art (image size of 384x288, label size of 16). 
As no other SotA ASIC chip supporting Bayes Net approximate inference is currently available, we benchmark our performance against running the same kernels on the multi-core RISC-V PULP platform.
Benefiting from our highly efficient sampler and parallel compiler algorithm, it shows superior performance across various Bayes Nets workloads and one MRF workload. Notably, we achieve an average 3.0x increase in throughput and an 2.8x improvement in energy efficiency on the different Bayesian Networks as shown in Fig. \ref{fig:application_result}. However, in scenarios where the cost of distribution generation is dominated, as seen in the "Art" workload, MSSE\cite{pgma} exhibits better performance. 

Table \ref{tab:full_area_power_breakdown} compares \name{} to the SotA, as the only implementation developed based on KY sampling algorithm that can accelerate any MCMC algorithm relying on discrete sampling, our work achieves up to 3.4x higher throughput and 2x higher area efficiency.

\begin{table}[!t]
\centering
\caption{Comparison table with SotA}
\begin{tabular}{lccccc}
\toprule
 & This work & MSSE\cite{pgma} & SPU\cite{spu} \\ 
 \midrule
Technology & 16nm & 16nm & FPGA  \\
Die Area & 4mm$^2$ & 5mm$^2$ & - \\
SRAM  & 960KB & 103KB & 4MB \\
SU No\# & 16 & 12 & 32  \\
$F_{max}$ & 300MHz & 651MHz & 146MHz   \\
Power@$F_{max}$ & 93.5mW@0.9V & 42.2mW@0.8V & -   \\
\makecell[l]{Peak Throughput \\ (GSample/s)} & 1.27 & 0.372$*$ & 4.67  \\
\makecell[l]{Peak Energy Effi. \\ (GSample/s/W)} & 20 & 17.6$*$ & -  \\
\makecell[l]{Peak Area Effi. \\ (GSample/s/mm$^2$)} & 0.567 & 0.284$*$ & -  \\
\midrule
Sampler & Knuth-Yao & CDF & CDF   \\ \midrule
Model & \makecell[c]{\textbf{General PMs} \\ (MRF, BayesNet)} & \makecell[c]{Only\\MRF} & \makecell[c]{Only\\MRF}   \\ \midrule
Inference & \makecell[c]{\textbf{discreate MCMC} \\ (Gibbs, MH, etc.)} & \makecell[c]{Only\\Gibbs} & \makecell[c]{Only\\Gibbs}   \\
\bottomrule
\end{tabular}%
\label{tab:full_area_power_breakdown}

\begin{tablenotes}
    \small
    \item $*$: contains both sampling and computing
\end{tablenotes}
\end{table}

\section*{Conclusion}
This work presents a 4mm$^2$ multicore SoC in 16nm, which executes approximate inference acceleration of PMs. It contains 4x4 customized RISC-V cores which feature non-normalized KY samplers and interpolation unit, as well as core-to-core direct data access via the register file. A customized compiler chain is developed to map MCMC algorithms for a wide variety of PMs. It shows the flexibility and performance gains compared to the SotA implementations.

\section*{Acknowledgment}
This project has been partly funded by the European Research Council (ERC) under grant agreement No. 101088865, Huawei, the Flanders AI Research Program and KU Leuven. Special thanks to ETH for the PULP platform support. 

\ifCLASSOPTIONcaptionsoff
  \newpage
\fi

\bibliographystyle{IEEEtran}
\bibliography{ref.bib}

\end{document}